\documentclass[a4paper, 11pt]{article}
\usepackage{framed}
\usepackage{graphicx}
\usepackage{xcolor}
\usepackage{slashed}
\usepackage{ amssymb }
\usepackage{mathtools}
\usepackage{latexsym}
\usepackage{textgreek}
\usepackage{verbatim}
 \usepackage[utf8]{inputenc}
 \usepackage{amsmath}
\usepackage{amsfonts}
\usepackage{verbatim}
\usepackage{graphicx}
\usepackage{subfigure}
\usepackage{amssymb}
\usepackage[T1]{fontenc}
\usepackage{slashed}
\usepackage{color}
\usepackage[numbers,sort&compress]{natbib}
\usepackage{amsmath}
\usepackage{mathrsfs}
\usepackage{physics}

\def\beq{\begin{eqnarray}}
\def\eeq{\end{eqnarray}}

\def\({\left(}
\def\){\right)}

\def\mpl{M_{\rm pl}}

\newcommand{\be}{\begin{equation}}
\newcommand{\ee}{\end{equation}}
\newcommand{\la}{\langle}
\newcommand{\ra}{\rangle}
\def\ea{\end{eqnarray}}
\def\ba{\begin{eqnarray}}

\def\beq{\begin{eqnarray}}
\def\eeq{\end{eqnarray}}

\def\({\left(}
\def\){\right)}
\def\mn{_{\mu \nu}}

\def\mpl{M_{\rm pl}}

\def\la{\langle}
\def\ra{\rangle}

\def\lsim{\mathrel{\rlap{\lower3pt\hbox{\hskip0pt$\sim$}}
     \raise1pt\hbox{$<$}}} 
\def\gsim{\mathrel{\rlap{\lower4pt\hbox{\hskip1pt$\sim$}}
     \raise1pt\hbox{$>$}}}

\def\lsim{\mathrel{\rlap{\lower3pt\hbox{\hskip0pt$\sim$}}
     \raise1pt\hbox{$<$}}} 
\def\gsim{\mathrel{\rlap{\lower4pt\hbox{\hskip1pt$\sim$}}
     \raise1pt\hbox{$>$}}}

\begin{document}

\begin{center}{\Large \bf{On Time-Evolution in Quantum Gravity
}}

 \vspace{1truecm}
\thispagestyle{empty} \centerline{\large  {Lasha  Berezhiani, Gia Dvali and Otari Sakhelashvili}}

\vskip 10pt

 \textit{Max-Planck-Institut f\"ur Physik, Werner–Heisenberg–Institut, 
 \\Boltzmannstra{\ss}e.~8, 85748 Garching, Germany
 \vskip 5pt
Arnold Sommerfeld Center, Ludwig-Maximilians-Universit\"at, \\Theresienstra{\ss}e 37, 80333 M\"unchen, Germany
 }

\end{center}  
 
\begin{abstract}

We derive an explicit BRST-exact operator identity for the bulk Hamiltonian in quantum gravity, working within a BRST-invariant quantization of General Relativity, treated as a low-energy effective field theory. We show that, up to a boundary term, the Hamiltonian can be written elegantly as the anticommutator of the BRST charge and the temporal ghost field. This form makes manifest that the Hamiltonian flow acts as a time-reparameterization on the correlation functions of the physical degrees of freedom. We demonstrate that the BRST-exactness of the bulk Hamiltonian does not trivialize the time evolution of gravitational backgrounds or bulk correlators, nor does it trivialize scattering amplitudes.

\end{abstract}

\newpage
\setcounter{page}{1}

\renewcommand{\thefootnote}{\arabic{footnote}}
\setcounter{footnote}{0}

\linespread{1.1}
\parskip 4pt

\section{Introduction} 

The concept of time-evolution has a special place within the quantum field theoretic description of our universe, which is founded on the Hamiltonian description of fundamental degrees of freedom. In quantum gravity, the challenges in the definition of time as the Hamiltonian flow have been widely discussed in the literature. This puzzling observation is connected to the vanishing of the classical Hamiltonian (modulo boundary contributions) upon satisfaction of constraint equations, which has dire consequences for the quantum theory. The question is how to sensibly quantize General Relativity (GR) while retaining time as the variable of the Hamiltonian flow. One of the criteria is the ability to accommodate the classical dynamics. Gravity being a gauge theory, one must take extra care in the identification of the physical Hilbert space preserved by the dynamics.

The important observations have been made in seminal papers \cite{DeWitt:1967yk,DeWitt:1967ub}, which sourced the discussion on this matter. On one hand, a certain attempt of canonical quantization of GR supplemented with the promotion of classical constraints into the physicality condition on the Hilbert space \cite{DeWitt:1967yk} seemed to yield an over-constrained Hamiltonian flow, with the boundary term of the Hamiltonian being the sole contributor to the time-dependence of physical states. This was argued to be somewhat undesirable as it did not make it transparent how one could recover the expected time-evolution of observables, consistent with classical gravity. On the other hand, the path-integral formulation of GR as the means for obtaining the scattering $S$-matrix elements seemed to be well-defined \cite{DeWitt:1967ub}. There, the question was left open as to whether a certain framework could consistently unify both of these avenues.

These questions are important for the identification of quantum states that serve as a fundamental description of classical backgrounds in gravity. The existence of time as the aforementioned variable of the Hamiltonian flow is essential for backgrounds corresponding to the time-dependent classical configurations.

In contrast, the issue of time does not appear in the computation of cosmological correlators for quantum fluctuations around the classical cosmological backgrounds. Therefore, if we hope to derive these results as an approximation of the background field method applied to a microscopic theory, the latter must possess a similar notion of time.

In \cite{Berezhiani:2024boz}, we have demonstrated that the BRST quantization of GR \cite{Kugo:1978rj} as a low energy effective field theory is ready-made for unifying the path-integral description of gravitational scattering amplitudes in Minkowski space-time and allowing for the well-defined notion of time as the parameter of the Hamiltonian flow for correlation functions under a single framework, reproducing classical dynamics in appropriate limit. The key observation of \cite{Berezhiani:2024boz} was that the quantum Hamiltonian and momentum constraints do not annihilate physical states within the quantization in question. As a result, physical states are not annihilated by the bulk Hamiltonian. 

In this work, we would like to further elucidate key aspects of this approach and confront the lore that there is no time in quantum gravity.

{The BRST-exactness of the bulk Hamiltonian is not claimed here as a new general theorem. In the Hamiltonian BRST literature, especially in the Batalin–Fradkin–Vilkovisky (BFV) formalism [6], this is the expected structural outcome for Hamiltonians built from first-class constraints. Our starting point, however, is not the BFV construction from the ungauge-fixed constrained system, but the gauge-fixed BRST Lagrangian formulation of General Relativity reviewed in Section 2. The purpose of the present work is to derive the corresponding canonical/operator realization and to explain, in that explicit framework, how the BRST-exact bulk Hamiltonian of General Relativity is compatible with nontrivial real-time evolution of bulk correlation functions and with the standard asymptotic dynamics once boundary contributions are included.}

\section{Framework}

Our starting point is the BRST-invariant Lagrangian formulation of GR \cite{Kugo:1978rj}
\beq
&&\mathcal{L}=\mathcal{L}_{\rm EH}+\mathcal{L}_{\rm GF}+\mathcal{L}_{\rm FP}\,,\\
&&\mathcal{L}_{\rm EH}=\sqrt{-g} \mpl^2R\,,\\
\label{gauge_fixing}
&&\mathcal{L}_{\rm GF}=\mpl b_\nu\partial_\mu\left(\sqrt{-g} g^{\mu\nu}\right)\,,\\
\label{ghostlagrangian}
&&\mathcal{L}_{\rm FP}=i \partial_\mu \bar{c}_\nu \left(\sqrt{-g}g^{\mu \sigma}\partial_\sigma c^{\nu}-\partial_\sigma\left(\sqrt{-g}g^{\sigma\nu}\right) c^{\mu}\right)\,.
\eeq
Here, in order to facilitate the quantization, the gauge has been fixed by supplementing the Einstein-Hilbert theory by $\mathcal{L}_{\rm GF}$, where  $b_\nu$ is the auxiliary field, the equation of motion for which imposes de Donder gauge on the metric
\beq
\partial_\mu\left(\sqrt{-g} g^{\mu\nu}\right)=0\,.
\eeq
Depending on the problem, a different gauge-fixing condition may be more convenient, particularly if one aims to recover the standard quantum field theory in curved spacetime around a specific background and in a chosen gauge (e.g. cosmological backgrounds in a particular slicing), as a limiting case of the background field method derived from our fundamental quantization approach \cite{Berezhiani:2024boz}; see also \cite{Berezhiani:2024pub} for related remarks on loop corrections to cosmological correlators. Moreover, for the realization of the BRST invariance, which is instrumental for the unitarity of the theory in the physical Hilbert space, the theory comes with the entourage of Faddeev-Popov ghost term $\mathcal{L}_{\rm FP}$ for fermionic ghost/anti-ghost fields. This contribution has been rewritten compared to \cite{Kugo:1978rj} by means of integration by parts to simplify the identification of canonical variables. 

The theory at hand enjoys the BRST symmetry, characterised by the Grassmannian parameter $\theta$. The transformation of the metric under this symmetry is identical to its properties under the coordinate reparameterizations with the parameter $(c^\mu\theta)$. The full set of fields at hand transforms as follows:
\beq
\label{dg}
&&\delta g\mn=-\mpl^{-1}\big( (\partial_\rho g\mn)c^\rho+g_{\mu\rho}\partial_\nu c^\rho+g_{\nu\rho}\partial_\mu c^\rho \big)\theta\,,\\
&&\delta c^\mu = \mpl^{-1}c^\rho\partial_\rho c^\mu\theta\,,\\
&&\delta \bar{c}_\mu=ib_\mu\theta\,,\\
&&\delta b_\mu=0\,.
\eeq

{The above transformations define the BRST differential $s$ by
$\delta\phi=(s\phi)\theta$ for any field $\phi$. Classically, this
differential is nilpotent:
\beq
s^2=0\,.
\eeq
The corresponding Noether charge $Q$ generates these
transformations. In canonical language, the same classical nilpotency
is equivalently expressed by the graded Poisson-bracket relation 
\beq
\{Q,Q\}=0\,.
\eeq
Upon quantization, this becomes the operator statement
\beq
\hat{Q}^2=0\,.
\eeq
and, together with the conserved ghost number, plays a central role in the identification of the physical Hilbert space over which the evolution/$S$-matrix is unitary.}

\section{Canonical Quantization}

At this point, the quantization is relatively straightforward, since all fields have canonical conjugate momenta. 
The special case being $b_\mu$, which virtually constitutes the conjugate momentum of the temporal metric components. {Since we are interested in the canonical/operator realization of the quantized theory and, in particular, in the emergence of real-time evolution for bulk correlation functions, we work in the explicit gauge-fixed BRST framework introduced in Section 2.} The canonical variables can be selected by convenience following \cite{Berezhiani:2024boz}:

\begin{itemize}

\item The spatial components of the metric are the obvious choice for characterising the on-shell propagating degrees of freedom, which we will denote by $\gamma_{ij}\equiv g_{ij}$. The corresponding canonical conjugate momentum is given by the well-known form
\beq
\Pi^{ij}=-\mpl^2\sqrt{\gamma}\left( K^{ij}-\gamma^{ij}K \right)\,,
\eeq
where $K_{ij}$ stands for the extrinsic curvature as usual. One of the advantages of writing the ghost Lagrangian in the above form is to maintain this classical expression for $\Pi^{ij}$.

\item The temporal components of the metric are most convenient to parameterize by means of $A^\mu\equiv \sqrt{-g}g^{\mu 0}$. The corresponding conjugate momentum reduces to
\beq
\Pi_\mu=\mpl b_\mu-i(\partial_\nu \bar{c}_\mu)c^\nu\,.
\eeq

\item The conjugate momenta for $\bar{c}_\mu$ and $c^\nu$ are
\beq
&&\Pi_{\bar{c}}^\mu=i\left( A^\sigma \partial_\sigma c^\mu-\partial_\sigma \left( \sqrt{-g} g^{\sigma\mu} \right)c^0 \right)\,,\\
&&\Pi_\nu^c=-iA^\mu\partial_\mu \bar{c}_\nu\,.
\eeq
\end{itemize}

In these canonical variables, the Einstein-Hilbert Hamiltonian takes the familiar form up to the boundary term
\beq
\label{EHH}
H_{\rm EH}=\int d^3 x \left[ -\frac{1}{A^0}\mathcal{H}_0+\frac{A^i}{A^0}\mathcal{P}_i \right]\,,
\eeq
where $\mathcal{H}_0$ and $\mathcal{P}_i$ are the Hamiltonian and momentum constraints respectively
\beq
\label{hamiltonianconst}
&&\mathcal{H}_0\equiv\frac{1}{2\mpl^2}\left( \gamma_{ik}\gamma_{j\ell}+\gamma_{i\ell}\gamma_{jk}- \gamma_{ij}\gamma_{k\ell}\right)\Pi^{ij}\Pi^{kl}-\mpl^2\gamma R^{(3)}\,,\\
&&\mathcal{P}_i\equiv-2\gamma_{ik}\partial_j\Pi^{kj}-(2 \partial_k\gamma_{ji}-\partial_i\gamma_{jk})\Pi^{jk}\,.
\label{momentumconst}
\eeq

This is obviously supplemented with the Hamiltonian for the auxiliary sector, which is given by
\beq
H_{FP+GF}=\int d^3 x \left( \frac{i}{A^0}\Pi^c_\mu\Pi^\mu_{\bar{c}}+\frac{A^k}{A^0}\left( \Pi_\mu^c \partial_k c^\mu-\partial_k\bar{c}_\mu\Pi^\mu_{\bar{c}}\right)+i\frac{1}{A^0}\gamma\gamma^{ij}\partial_i\bar{c}_\mu\partial_j c^\mu\right.\nonumber\\
\left.-\Pi_0\partial_k A^k-\Pi_i\partial_j\left(\frac{1}{A^0}\left(-\gamma\gamma^{ij}+A^i A^j\right)\right) \right)\,.
\eeq
In the formulation at hand, the constraint equations are promoted to the dynamical equations for auxiliary fields. In fact, we can use these equations to rewrite the Hamiltonian in a concise (yet non-canonical) form
\beq
H=H_{\rm EH}+H_{FP+GF}=\int d^3 x~A^\mu\partial_\mu\Pi_0\,.
\label{compH}
\eeq
This is connected to the fact that both the Hamiltonian and the momentum constraints appear in Hamilton's equation for $\Pi_0$. 

The quantization is straightforward and constitutes the promotion of the above-listed fields and conjugate momenta into operators, followed by the prescription of the canonical equal-time commutation/anti-commutation relations \cite{Berezhiani:2024boz}:
\beq
&&\left[\hat{\gamma}_{ij}(x),\hat{\Pi}^{k\ell}(y)\right]=\frac{i}{2}\left(\delta_{i}^{k}\delta_{j}^{\ell}+\delta_{i}^{\ell}\delta_{j}^{k}\right)\delta^{(3)}(x-y)\,,\\
&&\left[\hat{A}^\mu(x),\hat{\Pi}_\nu(y)\right]=i\delta^\mu_\nu\delta^{(3)}(x-y)\,,\\
&&\left\{\hat{c}^\mu(x),\hat{\Pi}^c_\nu(y)\right\}=i \delta^\mu_\nu\delta^{(3)}(x-y)\,,\\
&&\left\{\hat{\bar{c}}_\nu(x),\hat{\Pi}_{\bar{c}}^\mu(y)\right\}=i \delta^\mu_\nu\delta^{(3)}(x-y)\,.
\eeq
Henceforth, hats over quantities will indicate operators, while $[\ldots]$ and $\{\ldots\}$ will denote the commutators and anti-commutators respectively.

As a result of the straightforward computation, the Hamiltonian \eqref{compH} can be rewritten, up to a surface term, in terms of the BRST charge operator as
\beq
\hat{H}=\mpl\int d^3 x\left\{\hat{Q},i\hat{\Pi}^c_0\right\}\,.
\label{bulkham}
\eeq
This equality is the direct consequence of the BRST transformation property of $\hat{\Pi}^c_0$.
In other words, the bulk Hamiltonian, including the Einstein-Hilbert and ghost parts, is BRST exact. This is in perfect agreement with the general expectation that, for systems whose Hamiltonian is a linear combination of first-class constraints, it becomes BRST-exact up to BRST-invariant (non-exact) surface terms \cite{Henneaux:1994lbw}.
This is in a complete analogy of the world-line quantization of a relativistic particle \cite{VanHolten:2001nj}. The fact that the Hamiltonian in general relativity takes a BRST-exact form was previously identified in \cite{Ogawa:1997kq} {(see also \cite{Henneaux:2025ocw})}, albeit in a different parametrization and context, without drawing a connection to the concept of time as the Hamiltonian flow or lack thereof.
{The point of the present paper is therefore not the abstract statement of BRST-exactness itself, but its explicit realization in the adopted canonical variables and the interpretation of its consequences for real-time evolution of bulk correlators and, in the presence of boundaries, for asymptotic dynamics.}

In this work, we would like to focus on the effects of \eqref{bulkham}, which is the sole contributor to the evolution of the bulk correlation functions. In the presence of the boundary, one should keep in mind that there might be BRST-invariant surface contributions in addition, governing the evolution of the boundary correlators. In turn, the Hamiltonian \eqref{bulkham} tells the full story in the absence of the boundary.

The BRST invariance of the theory implies $[\hat{Q},\hat{H}]=0$, which is straightforwardly satisfied by \eqref{bulkham}. Furthermore, one of the classical features of General Relativity is that the Hamiltonian flow is equivalent to the reparametrization of time. In our framework, this is a direct consequence of \eqref{bulkham} and the fact that BRST symmetry acts on fields (with the exception of the auxiliary gauge-fixing sector) as the coordinate reparametrization with the parameter of the transformation given by the ghost fields; see e.g. \eqref{dg}. In fact, it is straightforward to show that for any operator $\hat{O}$ that commutes with $\hat{\Pi}^c_0$, we have
\beq
\label{dtO}
\partial_t \hat{O}(x)=i[\hat{H},\hat{O}(x)]=i\mpl\int d^3 z \left[\left\{\hat{Q},i\hat{\Pi}^c_0(z)\right\},\hat{O}(x)\right]=-i\mpl\int d^3 z \left\{\left[\hat{Q},\hat{O}(x)\right],i\hat{\Pi}^c_0(z)\right\}\,,
\eeq
where we have used the Jacobi identity in the last equality. Here, the commutator yields the BRST variation of $\hat{O}$, while the anti-commutator with $\hat{\Pi}_0$ projects the BRST transformation to the coordinate reparameterization in the time-direction. For instance, the verification of this statement for the infinitesimal evolution of $\hat{\gamma}_{ij}$ is straightforward. Namely, as a direct consequence of \eqref{dg}, we have
\beq
i\left[\hat{Q},\hat{\gamma}_{ij}\right]=\mpl^{-1}\Big( (\partial_\rho \hat{\gamma}_{ij})c^\rho+g_{i\rho}\partial_j c^\rho+g_{j\rho}\partial_i c^\rho \Big)\,.
\eeq
As a result
\beq
\left\{\left[\hat{Q},\hat{\gamma}_{ij}(x)\right],i\hat{\Pi}^c_0(z)\right\}=i\mpl^{-1}\Big( (\partial_0 \hat{\gamma}_{ij})\delta^{(3)}(x-z)+g_{i0}(x)\partial_j \delta^{(3)}(x-z)+g_{j0}(x)\partial_i \delta^{(3)}(x-z) \Big)\,,\nonumber
\eeq
where partial derivatives are with respect to $x$. Therefore, integration by $z$ would pass right through and nullify the last two terms, leaving us with
\beq
-i\mpl\int d^3 z\left\{\left[\hat{Q},\hat{\gamma}_{ij}(x)\right],i\hat{\Pi}^c_0(z)\right\}=\partial_t \hat{\gamma}_{ij}\,.
\eeq
This provides the explicit verification of \eqref{dtO} for the spatial metric.

\section{Physical States}

The full Hilbert space harbors unphysical states as well as physical ones, containing ghosts and unphysical polarizations of the gauge field. However, as it was thoroughly framed in \cite{Kugo:1979gm}, BRST quantization provides the mechanism for identifying the physical subspace with unitary $S$-matrix.

{In particular, physical states are required to have vanishing ghost number and to be annihilated by the BRST charge operator,
\beq
\hat{Q}|f\ra=0\,.
\label{physcond}
\eeq
This condition defines the kernel of $\hat Q$ in the zero-ghost-number sector, which we denote by $\mathcal{V}$. Within $\mathcal{V}$ there is the BRST-exact subspace, which we denote by $\mathcal V_0$, whose elements can be written as $\hat Q|\psi\ra$ for certain unphysical states $|\psi\ra$. Since $\mathcal{V}_0$ is the image of a linear operator, it is a subspace. Due to the nilpotency of $\hat{Q}$, the states in $\mathcal{V}_0$ have vanishing norm. The physical Hilbert space is identified with the quotient of BRST-closed states by BRST-exact states in the zero-ghost-number sector, i.e. with the BRST cohomology
\beq
\mathscr{H_{\rm phys}}\equiv \mathcal{V}/\mathcal{V}_0\,.
\label{quotient}
\eeq}

{In the absence of a boundary, eq. \eqref{bulkham} implies that the bulk Hamiltonian acts trivially on BRST cohomology. This does not mean, however, that chosen representatives of physical states or correlation functions of gauge-variant operators are time-independent. On the contrary, the invariance of cohomology classes under the bulk Hamiltonian flow is perfectly consistent with the nontrivial time dependence of their representatives and of the corresponding correlation functions. In fact, this is essential for recovering the property of GR that time evolution is equivalent to a reparameterization of time.}

To further elucidate this point, we will first consider the case of compactified spatial dimensions, before moving on to discuss the boundary Hamiltonian and its effects. We do so simply because the evolution of the bulk correlation functions is driven by the bulk Hamiltonian in any case.

\section{No-Boundary: Compact Space}

In this section, we focus on the case where the spatial manifold is closed, for instance, a three-sphere. We acknowledge the possible non-perturbative instabilities of such spacetimes \cite{Witten:1981gj}, but we will set these concerns aside and consider compact spatial dimensions for illustrative purposes. The primary aim of this discussion is to isolate the time dependence of correlation functions arising from the bulk Hamiltonian. We emphasize that our ultimate interest lies in spacetimes with Minkowski asymptotics.

In the absence of the boundary, the Hamiltonian of quantum gravity is fully given by \eqref{bulkham}. Which implies that the physical states are annihilated by the Hamiltonian up to an image (zero-norm state), i.e.
\beq
\label{hf}
\hat{H}|f\ra=\hat{Q}|\psi\ra\,,\qquad {\rm with} \qquad |\psi\ra=\mpl\int d^3 x~i\hat{\Pi}^c_0|f\ra\,.
\eeq
As far as $\mathscr{H}_{\rm phys}$ is concerned, $\hat{Q}|\psi\ra$ corresponds to its null element. This is precisely what one usually means when referring to the BRST-exact Hamiltonian as equivalent to zero. The repetitive application of the Hamiltonian yields a similar result, due to nilpotency $\hat{Q}^2=0$, resulting in
\beq
e^{-i\hat{H}t}|f\ra=|f\ra+\hat{Q}|\Psi\ra \,,
\label{imagetime}
\eeq
where $|\Psi\ra$ is yet another time-dependent unphysical state. The direct consequence of this observation is that the matrix element of the time evolution operator in physical states is time-independent
\beq
\label{f1Hf2}
\la f_1|e^{-i\hat{H}t}|f_2\ra=\la f_1|f_2\ra\,.
\eeq
In other words, the entire time-dependence of the state \eqref{imagetime} is the image and, consequently, the Hamiltonian flow does not change the elements of $\mathscr{H}_{\rm phys}$. 

To see, in what sense such a Hamiltonian flow has absolutely reasonable consequences, it suffices to examine the expectation value of any non-diffeomorphism invariant operator in the state \eqref{imagetime}. It is immediately clear that the image of the state, which captures the entire time-dynamics, gives a non-vanishing contribution to the matrix element, i.e.
\beq
\la f| \hat{O}|f\ra(t)=\la f| e^{i\hat{H}t}\hat{O}e^{-i\hat{H}t}|f\ra\neq\la f| \hat{O}|f\ra (t=0) \,.
\eeq
This follows from the fact that the image of \eqref{imagetime} does not generically have a vanishing contribution to the expectation value of gauge-variant operators, simply due to the physicality of $|f\ra$; e.g. $\la f|\hat{O}\hat{Q}|\Psi\ra\neq 0$. In order to illustrate the point, the latter matrix element can be rewritten in terms of the BRST variation as $\la f| {\rm i} \delta_{\scriptscriptstyle\rm B}\hat{O}|\Psi\ra$ in general. Taking into account that the variation of the gravitational field is linear in the ghost field \eqref{dg}, the matrix element in question to be non-vanishing, it is necessary for $|\Psi\ra$ to contain a state with a single ghost on top of the gravitational degrees of freedom, similar to $|\psi\ra$ from \eqref{hf}. All this is also reinforced by \eqref{dtO}, and in fact follows directly from it.

In summary, the physicality of the state, along with \eqref{f1Hf2}, does not preclude the time-evolution of correlation functions of reparametrization-non-invariant operators. 

\subsection{Nitty-Gritty of State Construction}

Having demonstrated that, let us discuss what type of BRST-invariant states would lead to the trivial time-evolution of all correlation functions, i.e., exhibiting features of DeWitt's states \cite{DeWitt:1967yk}. These are the states that at the first glance would have been most desirable to have due to their manifest BRST invariance, built by the action of an arbitrary BRST-invariant operator on a (BRST-invariant) Hamiltonian eigenstate, say a vacuum $|\Omega\ra$. {In the following discussion we assume that $|\Omega\ra$ is a Poincar\'e-invariant eigenstate of the Hamiltonian with vanishing energy, i.e. $\hat{H}|\Omega\ra=0$.}
In other words, as a direct consequence of \eqref{dtO}, we have
\beq
\hat{H}\hat{O}|\Omega\ra=0\,,
\eeq
for a (fermionic) bosonic operator $O$ that (anti-)commutes with $\hat{\Pi}^c_0$ and satisfies $[\hat{Q},\hat{O}]=0$.

Now let us ask, if not invariant, what should be the transformation property of ghost-independent operator $\hat{O}$ so that $\hat{O}|\Omega\ra$ is annihilated by $\hat{Q}$ but at the same time is not an eigenstate of $\hat{H}$. It is easy to see that
\beq
\hat{Q}\hat{O}|\Omega\ra=0\,, \qquad \text{as long as} \qquad [\hat{Q},\hat{O}]=\hat{\Delta}\hat{Q}\,,
\label{ocond}
\eeq
for an arbitrary $\hat{\Delta}$. Furthermore, according to \eqref{dtO}
\beq
[\hat{H},\hat{O}]\neq 0\,, \qquad\Longleftrightarrow \qquad \int d^3 z \left\{\hat{\Delta}\hat{Q},i\hat{\Pi}^c_0(z)\right\}\neq 0\,.
\label{dynamic_state}
\eeq
The important question is whether we can relax \eqref{ocond}.
The particular relaxation we are interested in concerns the introduction of nontrivial c-number functions in $\hat{O}$. Indeed, due to Poincar\'e invariance, the BRST invariance of the vacuum state $|\Omega\ra$ implies that it is also annihilated by the BRST charge density. In particular, denoting the latter by $\hat{\rho}_{\rm BRST}$, invariance of the vacuum can be written as
\beq
\int d^3 x e^{i\hat{P}\cdot x} \hat{\rho}_{\scriptscriptstyle BRST}(0)|\Omega\ra=0\,,
\eeq
where we have used the translation invariance of the vacuum. This opens up options to relax \eqref{ocond}, e.g. to a following possibility:
\beq
[\hat{Q},\hat{O}]=\int d^3 x\hat{\Delta}(x)\hat{\rho}_{\scriptscriptstyle BRST}(x)\,.
\label{chargedensitystate}
\eeq
Here, $\hat{\Delta}$ is once again an arbitrary operator that would depend on c-number functions defining $\hat{O}$. The explicit construction of such operators will be pursued elsewhere.

\section{Boundary: Asymptotic Minkowski}

Following the series of works \cite{Dvali:2011aa, Dvali:2012en, Dvali:2013eja, Dvali:2014gua, Berezhiani:2016grw, Dvali:2017eba, Berezhiani:2021zst, Berezhiani:2024boz, Dvali:2024dlb}, we envision our effective field theory formulation to be defined with the Minkowski vacuum, which is essential for the $S$-matrix formulation of the theory. Because of this, it is essential to extend the prior discussion to the asymptotically Minkowski spacetimes. Part of the discussion of the previous section changes significantly in the case of a non-compact space with a boundary. In particular, the total Hamiltonian acquires the boundary contribution, which is necessary for the consistency of the variation principle at the spatial boundary \cite{Regge:1974zd} and at the same time equates the total Hamiltonian to the Arnowitt--Deser--Misner (ADM) mass \cite{Arnowitt:1961zz}. The inclusion of the ADM surface term to the Hamiltonian in the context of quantum gravity goes back to \cite{DeWitt:1967yk}.

The need for additional surface contribution to \eqref{EHH}, for the consistency of the variation principle, is evident from the explicit form of the Hamiltonian constraint alone. In particular, a problematic contribution to the Einstein-Hilbert Hamiltonian can be cast in the form of the following surface integral
\beq
H_{\rm EH}\supset \mpl^2\int d^3 x ~\partial_k \left( \frac{1}{A^0}\gamma\gamma^{ij}\gamma^{k\ell}\left(\partial_i \gamma_{\ell j} -\partial_\ell\gamma_{ij}\right) \right)\,,
\label{surfcontr}
\eeq
which arises from the spatial Ricci scalar appearing in \eqref{hamiltonianconst}.

As it is well-known, the consistency requires the removal of such boundary contributions. This, in turn, will lead to the modification of \eqref{bulkham} by surface terms. {At this stage, we keep the boundary contribution schematic. Our purpose here is not to construct the full BRST-invariant surface term in generality, but only to emphasise that the presence of a boundary contribution changes the conclusion based on the BRST-exact bulk Hamiltonian alone.} So that the total Hamiltonian takes the following form
\beq
\hat{H}=\mpl\int d^3 x\left\{\hat{Q},i\hat{\Pi}^c_0\right\}+\int d^3x ~\partial_k\hat{\mathcal{A}}_k\,.
\label{Hboundary}
\eeq
In classical GR, it would have sufficed to require the variation of this surface contribution to cancel the variation of \eqref{surfcontr}. As a matter of fact, as we are focusing on asymptotically Minkowski spacetimes, $\mathcal{A}_k$ would be dictated by the linearized form of \eqref{surfcontr}; see \cite{Regge:1974zd} for the detailed discussion of the classical surface terms.

In the case of BRST quantization of GR, there are additional requirements that the surface term of \eqref{Hboundary} must satisfy. For starters, it must be BRST invariant. This immediately tells us a few important things. First and foremost, the matrix element of the Hamiltonian in physical states, and the expectation value as a special case among them, is given purely by the aforementioned surface term
\beq
\la f_1| \hat{H} |f_2\ra=\int dS^k\la f_1|\hat{\mathcal{A}}_k|f_2\ra \equiv \la f_1| {\hat E_{ADM}} |f_2\ra \,,
\eeq
where $dS^k$ denotes the surface element on the spatial boundary of the Minkowski space, and $E_{ADM}$ is the ADM mass/energy. {The corresponding linearized asymptotic expression is displayed later in \eqref{Eadm}.}

Furthermore, due to the BRST exactness of the bulk Hamiltonian, a similar relation holds for arbitrary powers of the Hamiltonian operator, which at the end of the day leads to the following physical matrix element of the time-evolution operator
\beq
\la f_1|e^{-i\hat{H}t}|f_2\ra=\la f_1|f_2\ra+\sum_{n=1}^\infty \frac{(-it)^n}{n!}\la f_1|\prod_{m=1}^n \int dS^{k}(z_m)\hat{\mathcal{A}}_k(z_m)|f_2\ra\,.
\label{scalarprod}
\eeq
Therefore, unlike \eqref{f1Hf2}, the matrix elements evolve in time, albeit purely due to boundary contribution. However, it is important to keep in mind that at the same time, the Hamiltonian flow of the correlation functions in the spatial bulk follows the BRST exact bulk Hamiltonian, just as it did in the no-boundary scenario. Generic states of physical relevance (e.g. coherent states) are expected to be wave packets, rather than eigenstates of the ADM energy \cite{DeWitt:1967yk}. Hence, they will possess nontrivial time evolution even without a bulk contribution. Providing a glimpse of the time-dependence of the scalar product \eqref{scalarprod}.

\section{Conserved Charges and the $S$-Matrix}

From the previous section, it is evident that we can define the energy associated with a physical state $\ket{f}$, for Minkowski asymptotics. 
 Irrespectively, whether the state is an eigenstate of the Hamiltonian or a coherent state, the expectation value of the energy will be conserved: 

\begin{equation}
\frac{d}{dt} E_{ADM} = \frac{d}{dt} \bra{f} \hat{H} \ket{f} = 0.
\end{equation}
This makes energy a good quantum number for the classification of physical states. Obviously, 
the physical states will be characterized by other quantum numbers. These were found in 
\cite{Regge:1974zd}, and they span the global Poincar\'e group, see e.g. \cite{Solovev:1985qdo},  and are represented by Casimirs and other conserved quantities which should be used to classify the physical Hilbert space, i.e., mass, spin, and energy-momentum { (for the recent developments see \cite{Henneaux:2025ocw})}. This means that we can define 
the asymptotic particle states and their interactions according to these quantum numbers.

The full theory can be specified by identifying the Hamiltonian $\hat{H}_0$ which defines the spectrum, and accounts for the rest in the form of interactions, see e.g. \cite{Weinberg:1995mt}. This is equivalent to finding the full $S$-matrix, which includes all elementary and bound/composite states. In practice, this is a difficult task, and we resort to perturbative methods. We approximate $H_0$ from the non-interacting limit of the theory, treating the rest as a perturbation.

At large distances, the perturbative description of GR simplifies to its quadratic form with the spectrum given by the graviton, which was studied in the BRST framework in \cite{Berezhiani:2021zst}. In the linear approximation, the charges take asymptotic forms as well, namely,
\begin{equation}
E_{ADM} = \mpl^2 \int d^3 x ~ \partial_{\ell} \left(\partial_i \gamma_{\ell i} - \partial_\ell \gamma_{ii} \right)=\int d^3x T_{00}\,,
\label{Eadm}
\end{equation}
where $T_{00}$ denotes the global time-translation Noether charge density for the linearized graviton.
The BRST charge $Q$ of the quadratic theory also tells us that the asymptotic states of the theory
 at low energies
contain only gravitons, and the non-linearities should be understood as their interactions, giving the usual $S$-matrix\footnote{In this framework, large black holes should be understood as 
intermediate 
multi-graviton resonances that eventually decay into the high multiplicity of asymptotic gravitons \cite{Dvali:2011aa}}.

We conclude this section with a brief discussion of a caveat related to the $S$-matrix in quantum gravity within the adopted framework. By construction, fully consistent physical states -- prior to truncating the perturbation theory at any given order -- are annihilated by the BRST charge. Consequently, the $S$-matrix elements defined via asymptotic states satisfying this condition cannot receive nontrivial contributions from the bulk Hamiltonian, for the reasons that led to \eqref{scalarprod}. At first glance, this may seem problematic for particle scattering, such as that of gravitons, since boundary terms typically do not contribute to scattering amplitudes of Fock states. However, BRST-invariant asymptotic states correspond to infrared-dressed states rather than undressed Fock states without soft graviton dressing. In this respect, the situation closely parallels that in quantum electrodynamics (see, e.g., \cite{Hirai:2019gio}). At tree level, however, such dressing is usually neglected because infrared divergences are absent at that order. Consequently, the corresponding asymptotic states are not fully BRST invariant, and their scattering amplitudes do receive contributions from the bulk Hamiltonian. Commonly, the $S$-matrix analysis is based on the Lagrangian path-integral formalism, which we demonstrated in \cite{Berezhiani:2024boz} to be equivalent to the Hamiltonian formalism within the BRST-invariant framework.

Thus, on Minkowski space, BRST quantization naturally restores the usual $S$-matrix and the classical equations of GR by consistency.

\section{Outlook}

{In this work, we have verified, within the gauge-fixed BRST framework adopted here, that the bulk Hamiltonian of General Relativity can be written in a BRST-exact form and have found its explicit compact expression. This is in agreement with the general Hamiltonian BRST expectation for systems whose Hamiltonian is built from first-class constraints. The point relevant for the present work is not the abstract existence of such a BRST-exact form, but its interpretation in the explicit canonical/operator framework and its implications for real-time evolution of bulk correlators and for asymptotic dynamics.}

To the best of our knowledge, Ref. \cite{Ogawa:1997kq} is the only reference where the full BRST-exactness of the Hamiltonian in canonical GR has been explicitly stated, i.e., expressing the total Hamiltonian entirely as a BRST anticommutator of the form $H=\int d^3 x\{Q,\Psi\}$, without residual Hamiltonian constraint terms. Although no discussion of consequences for the time-evolution as the Hamiltonian flow on the physical Hilbert space is given. The general discussion connecting the constraint part of the Hamiltonian, in addition to the auxiliary gauge-fixing sector, to the BRST-exact form has been given in \cite{Henneaux:1994lbw}. It has also been shown on a concrete simplified example that if the Hamiltonian consists of constraints alone, modulo gauge-fixing/Faddeev-Popov contributions, then it is fully BRST-exact and the Hamiltonian flow generates the time-reparameterization.

Therefore, the full Hamiltonian of quantum gravity with Minkowski vacuum reduces to this bulk term supplemented by the boundary contribution, corresponding to the ADM energy. We have, furthermore, demonstrated that the presence of this BRST-exact bulk contribution is vital in providing the time-dependent bulk correlation functions. Although the latter was already shown in \cite{Berezhiani:2024boz}, here we solidify the argument by linking it with this simplified form of the bulk Hamiltonian, in contrast to the common belief that the BRST-exact Hamiltonian is equivalent to the triviality of the Hamiltonian flow. We show that, although the cohomology of the BRST charge -- the space of equivalence classes that defines physical states -- is itself time-independent, particular representatives of those classes can carry time-dependence through their image. As a result, expectation values or correlation functions of gauge-variant operators (e.g. metric, curvature) acquire time-dependence expected from the semi-classical analysis. In fact, the quantum Hamiltonian acts as a time-reparameterization on all fields, with the exclusion of the auxiliary sector. Most importantly, our work strengthens the idea that the BRST-quantization of GR answers the puzzle posed by DeWitt \cite{DeWitt:1967ub}, where the author noticed the mismatch between seemingly over-restrictive canonical quantization and the path-integral formulation of quantum gravity. We even identify the possibility of the existence of the BRST-invariant states that would be analogues of DeWitt's physical states. In particular, the states constructed by the action of the BRST-invariant operators on the vacuum are automatically eigenstates of the bulk Hamiltonian, which was the entire Hilbert space in case of DeWitt's construction. We also pave the road for building a wider class of states within our framework.

Last but not least, we have highlighted how the nontrivial $S$-matrix of GR emerges despite the BRST-exactness of the bulk Hamiltonian.

\subsection*{Acknowledgments}

We would like to thank Giacomo Contri, Justin Khoury, and Archil Kobakhidze for invaluable discussions. The work of GD was supported in part by the Humboldt Foundation under the Humboldt Professorship Award, by the European Research Council Gravities Horizon Grant AO number: 850 173-6, by the Deutsche Forschungsgemeinschaft (DFG, German Research Foundation) under Germany’s Excellence Strategy - EXC-2111 - 390814868, Germany’s Excellence Strategy under Excellence Cluster Origins EXC 2094 – 390783311. The work of OS was partially supported by the Australian Research Council under the Discovery Projects grants DP210101636 and DP220101721.

\vskip 10pt

\noindent {\bf Disclaimer:} Funded by the European Union. Views and opinions expressed are, however, those of the authors only and do not necessarily reflect those of the European Union or European Research Council. Neither the European Union nor the granting authority can be held responsible for them.

\end{document}